\documentclass[12pt]{article}

\usepackage{amsmath}
\usepackage{amssymb}
\usepackage{amsfonts}
\usepackage{latexsym}
\usepackage{bm}
\usepackage{color}

\catcode `\@=11 \@addtoreset{equation}{section}

\catcode `\@=12

\newcommand{\be}{\begin{equation}}
\newcommand{\en}{\end{equation}}
\newcommand{\bea}{\begin{eqnarray}}
\newcommand{\ena}{\end{eqnarray}}
\newcommand{\beano}{\begin{eqnarray*}}
\newcommand{\enano}{\end{eqnarray*}}
\newcommand{\bee}{\begin{enumerate}}
\newcommand{\ene}{\end{enumerate}}

\newcommand{\mc}{\mathcal}

\newcommand{\D}{{\mc D}}

\newcommand{\Sc}{{\cal S}}

\newcommand{\F}{{\cal F}}
\newcommand{\G}{{\cal G}}

\newcommand{\Lc}{{\cal L}}

\newcommand{\1}{1 \!\! 1}

\newcommand{\Hil}{\mc H}

\newtheorem{thm}{Theorem}[section]

\newtheorem{defn}[thm]{Definition}

\catcode `\@=11 \@addtoreset{equation}{section}
\catcode `\@=12

\begin{document}

\thispagestyle{empty}

\vspace*{2cm}

\begin{center}
{{\Large \bf Appearances of pseudo-bosons from Black-Scholes equation}}\\[10mm]


{\large F. Bagarello} \footnote[1]{ Dipartimento di Energia, Ingegneria dell'Informazione e Modelli Matematici,
Facolt\`a di Ingegneria, Universit\`a di Palermo, I-90128  Palermo, and INFN, Universit\`a di di Torino, ITALY\\
e-mail: fabio.bagarello@unipa.it\,\,\,\, Home page: www.unipa.it/fabio.bagarello}


\end{center}

\vspace*{2cm}

\begin{abstract}
\noindent It is a well known fact that the Black-Scholes equation admits an alternative representation as a Schr\"odinger equation expressed in terms of a non self-adjoint hamiltonian. We show how {\em pseudo-bosons}, linear or not, naturally arise in this context, and how they can be used in the computation of the pricing kernel.

\end{abstract}


\vfill


\newpage

\section{Introduction}

Pseudo-Hermitian quantum mechanics, together with its many relatives, received an increasing interest by the physicists community and, more recently, by the mathematicians too, because of its possible applications in concrete systems and also in connection with several interesting mathematical properties appearing in systems driven by non-hermitian Hamiltonians. We refer to \cite{ben}-\cite{bagbook} for some recent and not so recent reviews or volumes dedicated to this subject.

Most examples of non hermitian Hamiltonians come from physics, see \cite{baginbagbook}-\cite{ellis} just to cite a few, or from mathematics, see \cite{bit2014}, where the interest is more focused to a rigorous treatment of the models under consideration, and to their main constituents.

But, as shown in \cite{baaquie} and, recently, in \cite{roy,roy2}, more examples of this kind of operators come from an unexpected field of research, i.e. from economics. In fact, with a suitable change of variable, the Black-Scholes equation can be rewritten as a Schr\"odinger equation, but with a non self-adjoint hamiltonian. This fact was used in \cite{baaquie}-\cite{roy2} to shown how quantum mechanical techniques can be useful to compute the pricing kernel for various situations, and how in some cases the Black-Scholes Hamiltonian can be factorized and used also as an interesting example in supersymmetric quantum mechanics, \cite{CKS}. In this paper we show that similar (or identical) models also provide  examples of $\D$-pseudo bosons ($\D$-PBs) and non linear $\D$-PBs ($\D$-NLPBs), which are excitations recently introduced by us, considering suitable deformations of the canonical commutation relations (CCRs). We also discuss how pricing kernels can be defined and computed in these cases.

The paper is organized as follows: in the next section we give a brief mathematical introduction to $\D$-PBs and $\D$-NLPBs, useful to keep the paper self-contained. In Section 3 we show how $\D$-PBs arise naturally out of the Black-Scholes equation, while in Section 4 we discuss the appearance of $\D$-NLPBs in the same context. Section 5 contains our conclusions.

\section{Mathematical preliminaries}

In view of their use in Sections 3 and 4, we devote this section to some preliminary definitions and results on $\D$-PBs and $\D$-NLPBs. We refer to  \cite{baginbagbook} and to \cite{bagnlpb1}-\cite{bagnlpb3} for more details.

\subsection{Something about $\D$-PBs}\label{sectpb}

Let $\Hil$ be a given Hilbert space with scalar product $\left<.,.\right>$ and related norm $\|.\|$. Let further $a$ and $b$ be two operators
on $\Hil$, with domains $D(a)$ and $D(b)$ respectively, $a^\dagger$ and $b^\dagger$ their adjoints, and let $\D$ be a dense subspace of $\Hil$
such that $a^\sharp\D\subseteq\D$ and $b^\sharp\D\subseteq\D$, where $x^\sharp$ is $x$ or $x^\dagger$. We are not requiring here that $\D$ coincides with, e.g. $D(a)$ or with $D(b)$. Nevertheless, for obvious reasons, $\D\subseteq D(a^\sharp)$
and $\D\subseteq D(b^\sharp)$.

\begin{defn}\label{def21}
The operators $(a,b)$ are $\D$-pseudo bosonic ($\D$-pb) if, for all $f\in\D$, we have
\be
a\,b\,f-b\,a\,f=f.
\label{21}\en
\end{defn}
 Sometimes, to simplify the notation, instead of (\ref{21}) we will simply write $[a,b]=\1$, having in mind that both sides of this equation
have to act on vectors of $\D$. Here $\1$ is the identity operator on $\Hil$.

\vspace{2mm}

Our  working assumptions are the following:

\vspace{2mm}

{\bf Assumption $\D$-pb 1.--}  there exists a non-zero $\varphi_{ 0}\in\D$ such that $a\,\varphi_{ 0}=0$.

\vspace{1mm}

{\bf Assumption $\D$-pb 2.--}  there exists a non-zero $\Psi_{ 0}\in\D$ such that $b^\dagger\,\Psi_{ 0}=0$.

\vspace{2mm}

Then, if $(a,b)$ satisfy Definition \ref{def21}, it is clear that $\varphi_0\in D^\infty(b):=\cap_{k\geq0}D(b^k)$ and that $\Psi_0\in D^\infty(a^\dagger)$, so
that the vectors \be \varphi_n:=\frac{1}{\sqrt{n!}}\,b^n\varphi_0,\qquad \Psi_n:=\frac{1}{\sqrt{n!}}\,{a^\dagger}^n\Psi_0, \label{22}\en
$n\geq0$, can be defined and they all belong to $\D$. Let $\F_\Psi=\{\Psi_{ n}, \,n\geq0\}$ and
$\F_\varphi=\{\varphi_{ n}, \,n\geq0\}$. Since each
$\varphi_n$ and each $\Psi_n$ belong to $\D$, they also belong to the domains of $a^\sharp$, $b^\sharp$, $N=ba$ and $N^\dagger=a^\dagger b^\dagger$. In particular, using (\ref{22}) and (\ref{21}) we deduce that
\be
\left\{
    \begin{array}{ll}
b\,\varphi_n=\sqrt{n+1}\,\varphi_{n+1}, \qquad\qquad\quad\,\, n\geq 0,\\
a\,\varphi_0=0,\quad a\varphi_n=\sqrt{n}\,\varphi_{n-1}, \qquad\,\, n\geq 1,\\
a^\dagger\Psi_n=\sqrt{n+1}\,\Psi_{n+1}, \qquad\qquad\quad\, n\geq 0,\\
b^\dagger\Psi_0=0,\quad b^\dagger\Psi_n=\sqrt{n}\,\Psi_{n-1}, \qquad n\geq 1,\\
       \end{array}
        \right.
\label{23}\en as well as the following eigenvalue equations: $N\varphi_n=n\varphi_n$ and  $N^\dagger\Psi_n=n\Psi_n$, $n\geq0$.  Then,  choosing the normalization of $\varphi_0$ and $\Psi_0$ in such a way $\left<\varphi_0,\Psi_0\right>=1$, we deduce that
\be \left<\varphi_n,\Psi_m\right>=\delta_{n,m}, \label{34a}\en
 for all $n, m\geq0$. This can be proved showing first that, if $n\neq m$, $\varphi_n$ and $\Psi_m$ are mutually orthogonal. This follows from the eigenvalues equations above for $N$ and $N^\dagger$. Then, using induction on $n$, one can check that $\left<\varphi_n,\Psi_n\right>=1$, for all $n\geq0$. We refer  to \cite{bagbook} and to \cite{bagnewpb} for more mathematical details on our framework, details which are mostly related to the fact that the operators involved in the game are almost all unbounded.

Let us now recall that a basis of $\Hil$ is a set of vectors $\F=\{f_n\in\Hil, \,n\geq0\}$, such that each vector $g\in\Hil$ can be written as $g=\sum_{n}c_nf_n$, with the complex coefficients $c_n$ uniquely determined. Here and in the following, convergence is always meant to be unconditional, in the norm topology of $\Hil$. Hence, our third assumption is the following:

\vspace{2mm}

{\bf Assumption $\D$-pb 3.--}  $\F_\varphi$ is a basis for $\Hil$.

\vspace{1mm}

This is equivalent to the request that $\F_\Psi$ is a basis for $\Hil$ as well, see for instance \cite{ole}, Theorem 3.3.2. In particular, if $\F_\varphi$ and $\F_\Psi$ are Riesz basis for $\Hil$, the $\D$-PBs are called {\em regular}.

\vspace{2mm}

In \cite{bagnewpb} a weaker version of Assumption $\D$-pb 3 has also been introduced for the first time, particularly interesting for physical applications:  let $\G$ be a suitable dense subspace of $\Hil$. Two biorthogonal sets $\F_\eta=\{\eta_n\in\G,\,n\geq0\}$ and $\F_\Phi=\{\Phi_n\in\G,\,n\geq0\}$ are called {\em $\G$-quasi bases} if, for all $f, g\in \G$, the following holds:
\be
\left<f,g\right>=\sum_{n\geq0}\left<f,\eta_n\right>\left<\Phi_n,g\right>=\sum_{n\geq0}\left<f,\Phi_n\right>\left<\eta_n,g\right>.
\label{25}
\en
Is is clear that, while Assumption $\D$-pb 3 implies (\ref{25}), the reverse is false, since this equation does not necessarily imply, for instance, that $g=\sum_{n\geq0}\left<\eta_n,g\right>\Phi_n$. However, if $\F_\eta$ and $\F_\Phi$ satisfy (\ref{25}), we still have some (weak) form of resolution of the identity\footnote{We recall that two biorthogonal sets $\F_f=\{f_n\in\Hil,\,n\geq0\}$ and $\F_g=\{g_n\in\Hil,\,n\geq0\}$ produce a {\em resolution of the identity} in $\Hil$ if, for all $f,g\in\Hil$, we get $\left<f,g\right>=\sum_{n\geq0}\left<f,f_n\right>\left<g_n,g\right>=\sum_{n\geq0}\left<f,g_n\right>\left<f_n,g\right>$. It is important to stress that, in (\ref{25}), $f$ and $g$ are assumed to belong to a dense subset of $\Hil$, and this is motivated by many physical models, \cite{baginbagbook}.}.  Now Assumption $\D$-pb 3 may be replaced by the following weaker condition:

\vspace{2mm}

{\bf Assumption $\D$-pbw 3.--}  $\F_\varphi$ and $\F_\Psi$ are $\G$-quasi bases for some subspace $\G$ dense in $\Hil$.

\vspace{2mm}

Let now Assumptions $\D$-pb 1,  $\D$-pb 2, and  $\D$-pbw 3 be satisfied. It might happen that $\G=\D$, but these sets can in general be different. Let us consider a self-adjoint, invertible, operator $\Theta$, which leaves, together with $\Theta^{-1}$, $\D$ invariant: $\Theta\D\subseteq\D$, $\Theta^{-1}\D\subseteq\D$. Then,  \cite{bagnewpb}, we say that $(a,b^\dagger)$ are $\Theta-$conjugate if $af=\Theta^{-1}b^\dagger\,\Theta\,f$, for all $f\in\D$. We can show that $(a,b^\dagger)$ are $\Theta-$conjugate if and only if  $(b,a^\dagger)$ are $\Theta-$conjugate. Moreover, one can prove that
 $(a,b^\dagger)$ are $\Theta-$conjugate if and only if $\Psi_n=\Theta\varphi_n$, for all $n\geq0$. Furthermore,
if $(a,b^\dagger)$ are $\Theta-$conjugate, then it turns out, \cite{baginbagbook}, that (i) $\left<f,\Theta f\right>>0$ for all non zero $f\in \D$, and that (ii) $\Theta Nf=N^\dagger \Theta f$, for all $f\in \D$. This last equality, which we can write as $\Theta N=N^\dagger \Theta $, is an intertwining relation, which is quite interesting in physics, see \cite{intop}, for instance.

\subsection{Something about $\D$-NLPBs}\label{sectnlpb}

$\D$-PBs have been proved to be quite useful to factorize and to find eigenvalues and eigenstates of several non self-adjoint Hamiltonians introduced along the years in the literature by many authors, Hamiltonians whose eigenvalues are linear in the quantum numbers of the system, see \cite{baginbagbook} for a (partial) list of applications. On the other hand, when this dependence is not linear, $\D$-PBs do not work well. For this reason, we have introduced a slightly extended definition of $\D$-PBs, which seems to provide a natural settings for these different situations: let us consider a strictly increasing sequence $\{\epsilon_n\}$: $0=\epsilon_0<\epsilon_1<\cdots<\epsilon_n<\cdots$. Then, given two
operators $a$ and $b$ on $\Hil$, and a set $\D\subset\Hil$ which is dense in $\Hil$, and which is stable under the action of $a^\sharp$ and $b^\sharp$, we introduce the following:

\begin{defn}\label{defnlpb}
We will say that the triple $(a,b,\{\epsilon_n\})$ is a family of  $ \D$-non linear pseudo-bosons ($\D$-NLPBs) if the following properties hold:
\begin{itemize}

\item {\bf p1.} a non zero vector $\Phi_0$ exists in $\D$ such that $a\,\Phi_0=0$;

\item {\bf  p2.} a non zero vector $\eta_0$ exists in $\D$ such that $b^\dagger\,\eta_0=0$;

\item {\bf { p3}.} Calling
\be \Phi_n:=\frac{1}{\sqrt{\epsilon_n!}}\,b^n\,\Phi_0,\qquad \eta_n:=\frac{1}{\sqrt{\epsilon_n!}}\,{a^\dagger}^n\,\eta_0, \label{55} \en we
have, for all $n\geq0$, \be a\,\Phi_n=\sqrt{\epsilon_n}\,\Phi_{n-1},\qquad b^\dagger\eta_n=\sqrt{\epsilon_n}\,\eta_{n-1}. \label{56}\en
\item {\bf { p4}.} The set $\F_\Phi=\{\Phi_n,\,n\geq0\}$ is a basis for $\Hil$.

\end{itemize}

\end{defn}

Notice that, since $\D$ is stable under the action of $b$ and $a^\dagger$, both $\Phi_n$ and  $\eta_n$ belong to $\D$, for all $n\geq0$. However, it might happen, and in fact it happens in Section 4, that $\Phi_n$ and $\eta_n$ belong to $\D$ {\em by themselves}, i.e. without any need to check if $\D$ is stable or not under the action of $a^\sharp$ and $b^\sharp$. This is important to have in mind, especially when this check is particularly complicated, as it happens in Section 4.  Incidentally, notice also that $\D$-PBs can be treated as a particular case of Definition \ref{defnlpb}, simply taking $\epsilon_n=n$.

As shown in \cite{bagnlpb1}, the set $\F_\eta=\{\eta_n,\,n\geq0\}$ is automatically a basis for $\Hil$ as well, and, calling $M=ba$, we deduce that
$M\Phi_n=\epsilon_n\Phi_n$ and $M^\dagger\eta_n=\epsilon_n\eta_n$. Therefore, choosing the normalization of $\eta_0$ and $\Phi_0$ in such a way
$\left<\eta_0,\Phi_0\right>=1$, $\F_\eta$ is biorthogonal to the basis $\F_\Phi$. This mean that $\F_\eta$ is the unique
basis which is biorthogonal to $\F_\Phi$.

\vspace{2mm}

As for $\D$-PBs, it could be reasonable to replace {\bf { p4}.} with some weaker requirement. In particular, it is natural to assume that

{\bf $\bullet$ { pw4}.} $\F_\Phi$ and $\F_\eta$ are $\G$-quasi bases for some $\G\subset\Hil$, dense in $\Hil$.

\vspace{2mm}

Also in this context it is possible to deduce interesting intertwining relations. For instance, if a self-adjoint, invertible and, in general, unbounded operator $\Theta$ exists which, together with $\Theta^{-1}$,
leaves $\D$ invariant, and such that $\eta_n=\Theta\,\Phi_n$, $\forall\,n$, see Section 2.1, then $\left(M^\dagger\Theta-\Theta M\right)\Phi_n=0,$ and $\left(\Theta^{-1} M^\dagger- M\Theta^{-1}\right)\eta_n=0$, for all $n=0,1,2,\ldots$.

\section{$\D$-PBs from the Black-Scholes equation}

After this brief mathematical introduction on $\D$-PBs and $\D$-NLPBs, we want to show how this general formalism appears in a rather natural way out of the Black-Scholes equation for option pricing with constant volatility $\sigma$,
\be
\frac{\partial C}{\partial t}=-\frac{1}{2}\sigma^2 S^2 \frac{\partial^2 C}{\partial S^2}-rS\frac{\partial C}{\partial S}-rC.
\label{31}
\en
Here $C(S,t)$ is the price of the option, $S$ is the stock price and $r$ is the risk-free spot interest rate. Introducing a new variable $x$ via $S=e^x$, and the related unknown function $\Psi(x)$ as $C(S(x),t)=e^{\epsilon t}\Psi(x)$, equation (\ref{31}) can be rewritten in the following form:
\be
H_{BS}\Psi(x)=\epsilon\Psi(x), \qquad \mbox{where}\qquad H_{BS}=-\frac{1}{2}\sigma^2  \frac{d^2 }{d x^2}+\left(\frac{\sigma^2}{2}-r\right)\frac{d}{dx}+r\,\1.
\label{32}\en
Notice that, even if (\ref{32}) looks like a Schr\"odinger equation, its Hamiltonian is not self-adjoint, $H_{BS}\neq H_{BS}^\dagger$, because of the presence of the term $\left(\frac{\sigma^2}{2}-r\right)\frac{d}{dx}$, which is linear in the $x$-derivative.

However, defining the multiplication operator by
\be
\rho=e^{-\beta x},\qquad \beta=\frac{1}{2}-\frac{r}{\sigma^2},
\label{33}
\en
it is possible to show that
\be
h_{BS}f(x):=\left(\rho H_{BS}\rho^{-1}\right)f(x)=\left[-\frac{1}{2}\sigma^2  \frac{d^2 }{d x^2}+\frac{1}{2\sigma^2}\left(\frac{\sigma^2}{2}+r\right)^2\1\right]f(x),
\label{34}\en
which holds for all $f(x)\in D(\Bbb R)$, the set of the $C^\infty$ functions with compact support. The reason for taking $f(x)$ in $D(\Bbb R)$ is because, on $D(\Bbb R)$, $\rho$ and $\rho^{-1}$ are surely well defined. More than this: $D(\Bbb R)$ is stable under the action of $\rho^{\pm\,1}$. In other words, $D(\Bbb R)\subseteq D(\rho)\cap D(\rho^{-1})$. However, it is clear that neither $\rho$ nor $\rho^{-1}$ are defined on all of $\Lc^2(\Bbb R)$: they are both unbounded operators.

For future convenience, rather than working on $D(\Bbb R)$, it is convenient to consider the set
$$
\D=\{f(x)\in \Sc(\Bbb R):\, e^{\gamma x}f(x)\in \Sc(\Bbb R),\, \forall\gamma\in{\Bbb C}\},
$$which contains $D(\Bbb R)$ as a subset, and for this reason is dense in $\Lc^2(\Bbb R)$. Here $\Sc(\Bbb R)$ is the set of all the $C^\infty$ functions which decay to zero, together with their derivatives, faster than any negative power. Equation (\ref{34}) also holds if $f(x)\in\D$. For brevity we will write

\be
h_{BS}=-\frac{1}{2}\sigma^2  \frac{d^2 }{d x^2}+\frac{1}{2\sigma^2}\left(\frac{\sigma^2}{2}+r\right)^2\1.
\label{34bis}\en
Notice that $h_{BS}=h_{BS}^\dagger$, since it is proportional to the square of the momentum operator $\hat p:=-i\frac{d}{dx}$, plus a real constant\footnote{We are adopting here the quantum mechanical notation, calling $\hat p$ the {\em momentum operator}. It is well known, \cite{messiah}, that both $\hat p$ and $\hat p^2=-\frac{d^2}{dx^2}$ are self-adjoint.}. In particular, $h_{BS}$ describes nothing but a free particle, of mass $\sigma^{-2}$, subjected to a constant potential. As for the Black-Scholes Hamiltonian $H_{BS}$, we find that
\be
H_{BS}^\dagger f(x)=\left(\rho^2 H_{BS}\rho^{-2}\right)f(x),
\label{35}\en
for all $f(x)\in \D$. As before, to simplify the notation, we will simply write $H_{BS}^\dagger =\rho^2 H_{BS}\rho^{-2}$: $\rho^2$ intertwines between $H_{BS}$ and $H_{BS}^\dagger$.

In \cite{baaquie,roy,roy2} some real potential $V(x)$ is added to the operator $H_{BS}$, defining in this way a new operator, $H_{eff}=H_{BS}+V(x)$. The economical interpretation of $V(x)$, see \cite{baaquie}, is that it can be used to represent a certain class of options; we will not insist on that here. Of course, $H_{eff}$ is also not self-adjoint, but can be easily mapped into a (formally) self-adjoint operator $h_{eff}=h_{eff}^\dagger$ extending what we have done above:
\be
h_{eff}:=\rho H_{eff}\rho^{-1}=h_{BS}+V(x), \quad \mbox{and}\quad H_{eff}^\dagger=\rho^2 H_{eff}\rho^{-2}.
\label{36}\en
These equalities could be made rigorous acting on functions of $\D$. In \cite{roy2} the authors' interest was focused on the factorization of $H_{eff}$, using standard results in supersymmetric quantum mechanics, \cite{CKS}. $H_{eff}$ can indeed be written as
\be
H_{eff}=BA+\delta\1,
\label{37}\en
where $\delta=\frac{\sigma^2\beta^2}{2}+r$,  $A$ and $B$ are operators defined as follows
\be
A=\frac{\sigma}{\sqrt{2}}\left(\frac{d}{dx}+W(x)-\beta\1\right), \quad
B=\frac{\sigma}{\sqrt{2}}\left(-\frac{d}{dx}+W(x)+\beta\1\right),
\label{38}\en
and where the real function $W(x)$ is related to the potential $V(x)$ as follows:
\be
V(x)=\frac{\sigma^2}{2}\left[W^2(x)-W'(x)\right].
\label{39}\en
With these definitions, in \cite{roy2} was also shown that $B=\rho^{-2}A^\dagger \rho^2$. Hence $B=A^\dagger$ if $\rho=\1$, which is true only if $\beta=0$, see (\ref{33}). This is clearly in agreement with the explicit expressions for $A$ and $B$ in (\ref{38}). Incidentally, in the language of Section 2.1, equality $B=\rho^{-2}A^\dagger \rho^2$ means that $(B,A^\dagger)$ are $\rho^2$-conjugate.

\subsection{How $\D$-PBs appear}

So far we are close to \cite{roy2}. Now, we focus on a different aspect of $H_{eff}$. In fact, rather than considering its supersymmetric partner, we look for the spectra of $H_{eff}$ and $H_{eff}^\dagger$ and for their eigenstates. For this reason, what is important for us, at least as a first step, is the commutator between $A$ and $B$ in (\ref{38}). A natural choice  consists in requiring that $[A,B]=\1$, the identity operator. This fixes the analytic form of $W(x)$, and of $V(x)$ as a consequence. In fact, since for all $f(x)\in \D$ we have $[A,B]f(x)=\sigma^2 W'(x)f(x)$, it is clear that $[A,B]=\1$ if
\be
W(x)=\frac{x}{\sigma^2}+w,
\label{310}\en
where $w$ is an arbitrary constant\footnote{From now on we will not write $\1$ explicitly, when there is no need.}, which we take real to ensure reality of $W(x)$ and, consequently, of $V(x)$. In fact, using (\ref{39}), we find that
\be
V(x)=\frac{\sigma^2}{2}\left(\frac{x}{\sigma^2}+w\right)^2-\frac{1}{2}.
\label{311}\en
This is a quadratic, real-valued, potential. Therefore, with this choice, $H_{eff}$ becomes a non self-adjoint shifted harmonic oscillator which appears, in slightly different ways, in several models in pseudo-hermitian quantum mechanics. We refer to \cite{baginbagbook}, and references therein, and to \cite{bgv}, for some examples of similar systems. Fixing $W(x)$ as in (\ref{310}) the operators $A$ and $B$ look like
\be
A=\frac{\sigma}{\sqrt{2}}\left(\frac{d}{dx}+\frac{x}{\sigma^2}+w-\beta\right), \quad
B=\frac{\sigma}{\sqrt{2}}\left(-\frac{d}{dx}+\frac{x}{\sigma^2}+w+\beta\right).
\label{312}\en
Moreover we have
\be
\left\{
    \begin{array}{ll}
H_{eff}=-\frac{1}{2}\sigma^2  \frac{d^2 }{d x^2}+
\frac{\sigma^2}{2}\left(\frac{x}{\sigma^2}+w\right)^2+\left(\frac{\sigma^2}{2}-r\right)\frac{d}{dx}+r-\frac{1}{2},\\
h_{eff}=-\frac{1}{2}\sigma^2  \frac{d^2 }{d x^2}+
\frac{\sigma^2}{2}\left(\frac{x}{\sigma^2}+w\right)^2+\frac{1}{2\sigma^2}
\left(\frac{\sigma^2}{2}+r\right)^2-\frac{1}{2}.\\
       \end{array}
        \right.
\label{313}\en
Recalling that $h_{eff}=\rho H_{eff} \rho^{-1}$, and that $H_{eff}=BA+\delta$, it is natural to introduce a new operator, $$c=\rho A\rho^{-1}=\frac{\sigma}{\sqrt{2}}\left(\frac{d}{dx}+\frac{x}{\sigma^2}+w\right),$$
again to be understood as an equality on $\D$. If we now compute $\rho B\rho^{-1}f(x)$, $f(x)\in \D$, we deduce that
$$
\rho B\rho^{-1}=\frac{\sigma}{\sqrt{2}}\left(-\frac{d}{dx}+\frac{x}{\sigma^2}+w\right)=c^\dagger.
$$
It is now possible to check that $h_{eff}$ can be written in terms of these operators as
$$h_{eff}=c^\dagger c+\delta,$$ and that $[c,c^\dagger]=\1$. Then, calling $\Phi_0(x)$ the vacuum of $c$, $c\,\Phi_0=0$, we find that, with a suitable choice of the normalization,
\be
\Phi_0(x)=\frac{1}{\pi^{1/4}\sqrt{\sigma}}\,e^{-\frac{1}{2}\left(\frac{x}{\sigma}+\sigma w\right)^2}.
\label{314}\en
Notice that $\|\Phi_0\|=1$ and that $\Phi_0(x)\in \Sc(\Bbb R)$ but also, more interesting for us, $\Phi_0(x)\in\D$. This function is an eigenstate of $h_{eff}$, with eigenvalue $E_0=\delta$. The other eigenstates of $h_{eff}$ can be constructed out of $\Phi_0(x)$ acting on this with powers of $c^\dagger$. We find:
\be
\Phi_n(x)=\frac{{c^\dagger}^n}{\sqrt{n!}}\Phi_0(x)=\frac{1}{\sqrt{\sigma\,2^n\,n!\sqrt{\pi}}}
\,H_n\left(\frac{x}{\sigma}+\sigma w\right)e^{-\frac{1}{2}\left(\frac{x}{\sigma}+\sigma w\right)^2},
\label{315}\en
for $n=0,1,2,\ldots$. Here $H_n(x)$ is the $n-$th Hermite polynomial. The set $\F_{\Phi}=\{\Phi_n(x),\, n\geq 0\}$ is an orthonormal basis for $\Lc^2(\Bbb R)$, and we have
\be
h_{eff}\Phi_n(x)=E_n\Phi_n(x),\qquad\mbox{where}\quad E_n=n+\delta
\label{316}\en
$n=0,1,2,3,\ldots$.

What we want to do now is to work directly with $A$ and $B$, and with $H_{eff}$ as in (\ref{37}), rather than with $c$ and its adjoint. The reason is that, as we have seen in several applications along the years, even if $H_{eff}$ is similar to $h_{eff}$, this does not imply that the properties of $h_{eff}$ and of its eigenstates automatically hold also for $H_{eff}$ and for its eigenvectors. In fact, the map which implements the similarity between these two operators is unbounded, and therefore not everywhere defined. This may have {\em bad} consequences, and in fact this is what we are now going to show.

\vspace{2mm}

Let $A$ and $B$ be the operators in (\ref{312}). As we have seen, $[A,B]f(x)=f(x)$, for all $f(x)\in \D$. $\D$ is stable under the action of $A$, $B$, and their adjoint. The solutions of $A\varphi_0=0$ and $B^\dagger\Psi_0=0$ satisfying $\left<\varphi_0,\Psi_0\right>=1$ can be easily found:
$$
\varphi_0(x)=N_\varphi\,e^{-\left(\frac{x^2}{2\sigma^2}+(w-\beta)x\right)},\quad \Psi_0(x)=N_\Psi\,e^{-\left(\frac{x^2}{2\sigma^2}+(w+\beta)x\right)},
$$
with $\overline{N_\varphi}\,N_\Psi=\frac{1}{\pi^{1/4}\sqrt{\sigma}}\,e^{\sigma^2w^2}$. Now, choosing $N_\varphi=N_\Psi$ real, we find that
\be
\left\{
    \begin{array}{ll}
\varphi_0(x)=e^{\beta x}\Phi_0(x)=\rho^{-1}\,\Phi_0(x),\\
\Psi_0(x)=e^{-\beta x}\Phi_0(x)=\rho\,\Phi_0(x).\\
       \end{array}
        \right.
\label{314b}\en
It is clear that neither $\varphi_0(x)$ nor $\Psi_0(x)$ belong to $D(\Bbb R)$, while they both belong to $\D$. This explains why $\D$ is more relevant for us than the sets $D(\Bbb R)$. $\D$ is also more useful than $\Sc(\Bbb R)$, because, even if $\varphi_0(x), \Psi_0(x)\in\Sc(\Bbb R)$, $\Sc(\Bbb R)$ is not stable under the action  of $\rho^{\pm \,1}$, which is something useful for us. Incidently, we also observe that each $\Phi_n(x)$ in (\ref{315}) belongs to $\Sc(\Bbb R)$ and to $\D$, but not to $D(\Bbb R)$.

By induction we can now check that (\ref{314b}) extends to all the other functions, i.e. that, defining $\varphi_n(x)$ and $\Psi_n(x)$ as in (\ref{22}), then
\be
\left\{
    \begin{array}{ll}
\varphi_n(x)=\rho^{-1}\,\Phi_n(x),\\
\Psi_n(x)= \rho\,\Phi_n(x),\\
       \end{array}
        \right.
\label{315b}\en
for $n=0,1,2,\ldots$. Here $\Phi_n(x)$ are given in (\ref{315}). Of course, each $\varphi_n(x)$ and each $\Psi_n(x)$ belong to $\D$. It might be interesting to observe that $\Psi_n(x)$ can be deduced directly from $\varphi_n(x)$ simply replacing $\beta$ with $-\beta$.  Now, let us consider the sets $\F_\Psi=\{\Psi_{ n}, \,n\geq0\}$ and
$\F_\varphi=\{\varphi_{ n}, \,n\geq0\}$. These are two biorthogonal sets of eigenvectors of $N=BA$ and $N^\dagger=A^\dagger B^\dagger$:
$$
N\varphi_n(x)=n\varphi_n(x),\quad N^\dagger\Psi_n(x)=n\Psi_n(x),\quad \left<\varphi_n,\Psi_m\right>=\delta_{n,m}.
$$
Our previous results show that, if $\beta=0$ (i.e. if $\sigma^2=2r$), then $\varphi_n(x)=\Psi_n(x)=\Phi_n(x)$ for all $n$. Hence the sets $\F_\varphi$ and $\F_\Psi$ both coincide with $\F_\Phi$, which is an orthonormal basis for $\Hil=\Lc^2(\Bbb R)$. Hence a natural question is whether $\F_\varphi$ and $\F_\Psi$ are also bases if $\beta\neq0$.
To prove that, in this case, they are indeed {\bf not} bases, we will compute first $\|\varphi_n\|$ and $\|\Psi_n\|$. This can be done using Formula 7.374.7 of \cite{grad}:
$$
\int_{\Bbb R}e^{-(x-y)^2}H_m(x)H_n(x)dx=2^n\sqrt{\pi}\,m!\,y^{n-m}L_m^{n-m}(-2y^2),
$$
if $m\leq n$. Here $L_m^{k}$ is a Laguerre polynomial. Hence,
$$
\|\varphi_n\|^2=e^{\beta^2\sigma^2-2\beta w \sigma^2}\,L_n(-2\beta^2\sigma^2), \quad \|\Psi_n\|^2=e^{\beta^2\sigma^2+2\beta w \sigma^2}\,L_n(-2\beta^2\sigma^2),
$$
which are both diverging if $n\rightarrow\infty$, when $\beta\neq0$, see \cite{szego}, Theorem 8.22.3. We can now use \cite{dav}, Lemma 3.3.3, to prove that $\F_\varphi$ and $\F_\Psi$ are indeed not bases:

let us assume for a moment that $\F_\varphi$ is a basis for $\Hil$. Hence each $f\in\Hil$ can be written as
$f=\sum_{n=0}^\infty\left<\Psi_n,f\right>\varphi_n=\sum_{n=0}^\infty P_n(f)$, where $P_n(f):=\left<\Psi_n,f\right>\varphi_n$. However, since $\|P_n\|=\|\varphi_n\|\|\Psi_n\|$,  and since $\|P_n\|=\|\varphi_n\|\|\Psi_n\|\rightarrow\infty$,  $\sup_n\|P_n\|=\infty$. As a consequence, the above
expansion cannot converge for all vectors $f\in\Hil$. Hence, $\F_\varphi$ cannot be a basis for $\Hil$. In a similar way we can conclude that
$\F_\Psi$ cannot be a basis for $\Hil$. On the other hand, since when $\beta=0$ the argument of the Laguerre functions $L_n$ is zero, and since $L_n(0)=1$ for all $n$, we deduce that $\|\varphi_n\|= \|\Psi_n\|=1$. Hence $\sup_n\|P_n\|=1$, and the argument above does not apply. This is in agreement with what we have seen before: if $\beta=0$, $\F_\varphi$ and $\F_\Psi$ coincide and they give rise to a (single) orthonormal basis of $\Hil$.

Going back to the case $\beta\neq 0$, it is still possible to show that $\F_\varphi$ and $\F_\Psi$ are $\D$-quasi bases. For that, it is sufficient to recall that $\D$ is stable under the action of $\rho$ and $\rho^{-1}$. Hence, taken $f,g\in\D$, we have
$$
\left<f,g\right>=\left<\rho f,\rho^{-1}g\right>=\sum_{n=0}^\infty\left<\rho f,\Phi_n\right>\left<\Phi_n,\rho^{-1}g\right>=
\sum_{n=0}^\infty\left< f,\varphi_n\right>\left<\Psi_n,g\right>,
$$
and analogously $\left<f,g\right>=
\sum_{n=0}^\infty\left< f,\Psi_n\right>\left<\varphi_n,g\right>$. In the derivation we have used the fact that $\F_\Phi$ is an orthonormal basis for $\Hil$.

\vspace{2mm}

Other results which can be easily deduced are the following:

\begin{enumerate}

\item $(A,B^\dagger)$ are $\Theta$-conjugate, where $\Theta:=\rho^2=e^{-2\beta x}$. This can be checked computing explicitly $Af$ and $\Theta^{-1}B^\dagger \Theta f$, for $f\in\D$, and noticing that they coincide.

\item $\Psi_n(x)=\Theta\varphi_n(x)$, for all $n=0,1,2,\ldots$.

\item $\Theta$ is a positive operator.

\end{enumerate}

Then the general framework discussed in Section \ref{sectpb} is recovered in all its aspects. Of course, we also have
\be
H_{eff}\varphi_n=E_n\varphi_n,\qquad H_{eff}^\dagger\Psi_n=E_n\Psi_n,
\label{317}
\en
for all $n=0,1,2,\ldots$: $H_{eff}$ and $H_{eff}^\dagger$ are isospectral, as expected, and they are also isospectral to $h_{eff}$, see (\ref{316}). In fact, the following intertwining relations can be easily deduced:
$$
\rho H_{eff}=h_{eff}\rho,\quad H_{eff}^\dagger\rho=\rho h_{eff},\quad\mbox{and }\quad \Theta H_{eff}=H_{eff}^\dagger\Theta,
$$
 all well defined on $\D$.

Going back to the economic aspect of the model, following \cite{baaquie,roy,roy2} the next step consists in the computation of the price kernel, $\left<x,e^{-\tau H}x'\right>$. Here $\left|x\right>$ is the generalized eigenstate of the position operator written, in agreement with \cite{baaquie,roy,roy2}, as a Dirac bra.

 In the present context, due to the existence of two equally relevant Hamiltonians, $H_{eff}$ and $H_{eff}^\dagger$, we have two possible choices for the price kernel:
$$
p_1(x,x';\tau)=\left<x,e^{-\tau H_{eff}}x'\right>, \qquad p_2(x,x';\tau)=\left<x,e^{-\tau H_{eff}^\dagger}x'\right>.
$$
Assuming that $\F_\varphi$ and $\F_\Psi$ produce a resolution of the identity\footnote{This is not granted by the fact that $\F_\varphi$ and $\F_\Psi$ are $\D$-quasi bases, since  neither $\left|x\right>$ nor $e^{-\tau H}\left|x'\right>$ belong to $\D$, obviously. However, $\D$ is not necessarily the biggest set on which $\F_\varphi$ and $\F_\Psi$ produce a resolution of the identity.}, we have
$$
p_1(x,x';\tau)=\sum_{n=0}^\infty \left<x,e^{-\tau H_{eff}}\varphi_n\right>\left<\Psi_n, x'\right>=e^{-\tau\delta+\beta(x-x')}\sum_{n=0}^\infty e^{-\tau n}\Phi_n(x)\overline{\Phi_n(x')},
$$
where we have used (\ref{315b}) and (\ref{317}). Recalling the explicit form of the function $\Phi_n(x)$ in (\ref{315}), and using the following formula for the Hermite polynomials,
$$
(1-z^2)^{-1/2}\exp\left\{y^2-\frac{(y-zx)^2}{1-z^2}\right\}=\sum_{n=0}^\infty \frac{1}{n!}H_n(x)H_n(y)\left(\frac{z}{2}\right)^n
$$
see \cite{magnus}, pg. 252, we find that
\be
p_1(x,x';\tau)=\frac{1}{\sigma\sqrt{\pi}}\,e^{-\tau\delta+\beta(x-x')}e^{-\frac{1}{2}\left[
\left(\frac{x}{\sigma}+\sigma w\right)^2+\left(\frac{x'}{\sigma}+\sigma w\right)^2\right]}I(x,x';\tau),
\label{318}
\en
where

\be
I(x,x';\tau)=\frac{1}{\sqrt{1-e^{-2\tau}}}
\exp\left\{\left(\frac{x'}{\sigma}+\sigma w\right)^2-
\frac{1}{1-e^{-2\tau}}\left(\frac{x'}{\sigma}+\sigma w-e^{-\tau}\left(\frac{x}{\sigma}+\sigma w\right)\right)^2\right\}.
\label{319}\en

\vspace{2mm}

{\bf Remarks:--} (1) It might be interesting to observe that $I(x,x';\tau)$ is well defined when $\tau\neq0$. This is in agreement with the fact that $p_1(x,x';0)=\delta(x-x')$, see \cite{baaquie,roy,roy2}.

(2) Concerning $p_2(x,x';\tau)$, not many differences arise. The only point is that, when using the resolution of the identity in its definition (assuming again that this holds true) we have to exchange the roles of  $\varphi_n$ and $\Psi_n$, to use the fact that each $\Psi_n$ is an eigenstate of $H_{eff}^\dagger$ (while $\varphi_n$ is not!):
$$
p_2(x,x';\tau)=\sum_{n=0}^\infty \left<x,e^{-\tau H_{eff}^\dagger}\Psi_n\right>\left<\varphi_n, x'\right>.
$$
Similar computations show that the result differs from $p_1(x,x';\tau)$ just for $\beta$ being replaced by $-\beta$, as expected.

\section{$\D$-NLPBs from the Black-Scholes equation}

In this section, more on the lines of \cite{baaquie} and \cite{roy}, we consider a different choice of $V(x)$ in $H_{eff}$, and we show how the general structure sketched in Section \ref{sectnlpb} is recovered in this case.

The potential $V(x)$ we consider here is no longer the one in (\ref{39}). $V(x)$ is now a double knock out barrier defined as
$$
V(x)=\left\{
    \begin{array}{ll}
0\qquad\,\, \mbox{ if } x\in]a,b[\\
\infty,\qquad \mbox{otherwise}.\\
       \end{array}
        \right.
$$
As before we have $h_{eff}=\rho H_{eff}\rho^{-1}=h_{BS}+V(x)$, see (\ref{36}), where $h_{BS}$ is given in (\ref{34bis}), which we rewrite here as
\be
h_{BS}=-\frac{1}{2}\sigma^2  \frac{d^2 }{d x^2}+\gamma,\qquad \gamma:=\frac{1}{2\sigma^2}\left(\frac{\sigma^2}{2}+r\right)^2
\label{41}\en
The eigensystem for $h_{eff}$ is well known in the physical literature. We have
\be
\left\{
    \begin{array}{ll}
h_{eff}\Phi_n(x)=\tilde\epsilon_n\Phi_n(x),\qquad n=0,1,2,\ldots,\\
\Phi_n(x)=\sqrt{\frac{2}{b-a}}\,\sin\left[\lambda_{n+1}(x-a)\right],\\
\lambda_n=\frac{n\pi}{b-a},\qquad\qquad\quad\qquad \tilde\epsilon_n=\frac{\sigma^2\lambda_{n+1}^2}{2}+\gamma.\\
       \end{array}
        \right.
\label{42}\en

From a mathematical point of view the situation is, in a sense, slightly simpler than the one in the previous section, since the Hilbert space of this model is $\Hil_1=\Lc^2(a,b)$, and both $\rho$ and $\rho^{-1}$ turn out to be bounded operators on $\Hil_1$. Then, defining $\varphi_n(x)=\rho^{-1}\,\Phi_n(x)$ and $\Psi_n(x)= \rho\,\Phi_n(x)$ as in (\ref{315b}), they are automatically well defined in $\Hil_1$, and biorthogonal: $\left<\varphi_n,\Psi_m\right>_1=\delta_{n,m}$, where $\left<.,.\right>_1$ is the scalar product in $\Lc^2(a,b)$. Moreover,
\be
H_{eff}\,\varphi_n(x)=\tilde\epsilon_n\,\varphi_n(x),\qquad H_{eff}^\dagger\,\Psi_n(x)=\tilde\epsilon_n\,\Psi_n(x),
\label{43}\en
$n=0,1,2,\ldots$. Now, since $\rho^{\pm 1}$ are bounded, and since $\F_{\Phi}=\{\Phi_n(x), n\geq0\}$ is an orthonormal basis for $\Hil_1$, $\F_\varphi=\{\varphi_n(x), n\geq0\}$ and $\F_\Psi=\{\Psi_n(x), n\geq0\}$ are Riesz bases for $\Hil_1$.

In \cite{roy} $H_{eff}$ is shown to be (essentially) factorizable. In fact, introducing
\be
A=\frac{d}{dx}-\lambda_1\cot\left[\lambda_1(x-a)\right]-\beta,\qquad B=-\frac{d}{dx}-\lambda_1\cot\left[\lambda_1(x-a)\right]+\beta,
\label{44m}\en
we have
\be
H_{eff}\,\varphi_n(x)=\left[\frac{\sigma^2}{2}BA+\left(\gamma+\frac{\sigma^2\lambda_1^2}{2}\right)
\right]\varphi_n(x)=\tilde\epsilon_n\varphi_n(x),
\label{44}\en
for all $n=0,1,2,\ldots$. We see that $A\neq B^\dagger$ if $\beta\neq 0$, which is in agreement with our previous analysis of $H_{eff}$.

\subsection{This factorization does not work}

We extend (\ref{44}) to the linear span of the $\varphi_n$'s, which, being $\F_\varphi$ a Riesz basis, is dense in $\Hil_1$. Briefly we write
$$
H_{eff}=\frac{\sigma^2}{2}BA+\delta,\qquad \delta=\gamma+\frac{\sigma^2\lambda_1^2}{2}.
$$
Also, $H_{eff}^\dagger=\frac{\sigma^2}{2}A^\dagger B^\dagger+\delta$. In the attempt to use in the present context the general settings discussed in Section \ref{sectnlpb}, we have first to control if $A\varphi_0=0$ and $B^\dagger\Psi_0=0$. With our definitions these are both true, as one can explicitly check. Hence {\bf p1} and {\bf p2} are satisfied. However, it is also easy to see that {\bf p3} is not satisfied, since, in particular, $B\varphi_n$ is not proportional to $\varphi_{n+1}$. In fact,  while $B\varphi_0(x)$ is proportional to $e^{\beta x}\cos(\lambda_1(x-a))$, $\varphi_1(x)$ is proportional to $e^{\beta x}\sin(2\lambda_1(x-a))$: hence $B\varphi_0(x)$ and $\varphi_1(x)$ are different.

The conclusion is therefore that the operators $A$ and $B$ in (\ref{44m}), even if they are useful to factorize $H_{eff}$, cannot give rise to a family of $\D$-NLPBs as defined in Section \ref{sectnlpb}. However, this does not exclude that different factorizations do exist which make this possible, and in fact this is what we are going to show next.

\subsection{A different factorization}

We start our analysis noticing that, if $H_{eff}$ has to be factorized in terms of a different pair of operators $\hat A$ and $\hat B$, the following equality should be true:
\be
\left[\frac{\sigma^2}{2}\hat B\hat A+\delta
\right]\varphi_n(x)=\tilde\epsilon_n\varphi_n(x),
\label{45m}\en
so that
\be
\hat B\hat A\varphi_n(x)=\rho_n\varphi_n(x),
\label{45}\en
where $\rho_n=\lambda_{n+1}^2-\lambda_1^2=\frac{\pi^2 n(n+2)}{(b-a)^2}$.

In our next construction, we will make use of the fact that $\F_\varphi$ and $\F_\Psi$ are biorthogonal Riesz bases for $\Hil_1$, and we will adopt here the general framework originally proposed, in a rather abstract version, in \cite{bit2014}. This is natural, since the main mathematical tool used in \cite{bit2014} is exactly a pair of biorthogonal Riesz bases, as the one we have here.

We first introduce the operators $\hat B$ and $\hat A$ as follows:
$$
D(\hat B)=\left\{f\in\Hil_1: \sum_{n=0}^\infty \sqrt{\rho_{n+1}}\, \left<\Psi_n,f\right>_1\varphi_{n+1} \,\mbox{ exists in }\Hil_1\right\},
$$
$$
D(\hat A)=\left\{f\in\Hil_1: \sum_{n=1}^\infty \sqrt{\rho_{n}}\, \left<\Psi_n,f\right>_1\varphi_{n-1} \,\mbox{ exists in }\Hil_1\right\},
$$
and
\be
\hat B f=\sum_{n=0}^\infty \sqrt{\rho_{n+1}}\, \left<\Psi_n,f\right>_1\varphi_{n+1},\qquad \hat A g=\sum_{n=1}^\infty \sqrt{\rho_{n}}\, \left<\Psi_n,g\right>_1\varphi_{n-1},
\label{46}\en
for all $f\in D(\hat B)$ and $g\in D(\hat A)$. These two operators are densely defined since, calling $\Lc_\varphi$  the linear span of the $\varphi_n$'s, which is dense in $\Hil_1$, we have $\Lc_\varphi\subseteq D(\hat B)$ and $\Lc_\varphi\subseteq D(\hat A)$. Following \cite{bit2014} we can rewrite these domains as
$$
D(\hat B)=\left\{f\in\Hil_1: \sum_{n=0}^\infty \rho_{n+1}\, |\left<\Psi_n,f\right>_1|^2<\infty \right\}, \, D(\hat A)=\left\{g\in\Hil_1: \sum_{n=0}^\infty \rho_{n}\, |\left<\Psi_n,g\right>_1|^2<\infty \right\}.
$$
Moreover, see \cite{bit2014}, we also have
\be
\hat B\hat A f=\sum_{n=0}^\infty \rho_{n}\, \left<\Psi_n,f\right>_1\varphi_{n}, \quad
\hat A\hat B g=\sum_{n=0}^\infty \rho_{n+1}\, \left<\Psi_n,g\right>_1\varphi_{n},
\label{47}\en
where $f\in D(\hat B\hat A)$ and $g\in D(\hat A\hat B)$, with
$$
D(\hat B\hat A)=\left\{f\in\Hil_1: \sum_{n=0}^\infty \rho_{n}^2\, |\left<\Psi_n,f\right>_1|^2<\infty \right\}, \, D(\hat A\hat B)=\left\{g\in\Hil_1: \sum_{n=0}^\infty \rho_{n+1}^2\, |\left<\Psi_n,g\right>_1|^2<\infty \right\}.
$$
Both $D(\hat B\hat A)$ and $D(\hat A\hat B)$ are dense in $\Hil_1$, since $\Lc_\varphi$ is contained in both these sets. A direct computation now shows that, as expected, $\varphi_n\in D(\hat B\hat A)$ and $\hat B\hat A\varphi_n=\rho_n\varphi_n$, $\forall\,n$, see (\ref{45}), and $\hat A\hat B\varphi_n=\rho_{n+1}\varphi_n$, $\forall\,n$.

\vspace{2mm}

Our next goal is to show that $\left(\hat A,\hat B,\{\rho_n\}\right)$ is a family of $\hat\D$-NLPBs, for $\hat D$ defined as follows:
$$
\hat D=\left\{f(x)\in\Lc^2(a,b):\, f'(x)\in\Lc^2(a,b)\right\}.
$$
This set is dense in $\Lc^2(a,b)$, since it contains the linear span of the functions $h_k(x):=\frac{1}{\sqrt{b-a}}\,e^{i\frac{2\pi k x}{b-a}}$, $k\in {\Bbb Z}$, which is dense in $\Hil_1$. In fact, $\{h_k(x),\,k\in {\Bbb Z}\}$ is an orthonormal basis for $\Hil_1$.

The functions $\varphi_0(x)=\sqrt{\frac{2}{b-a}}\,e^{\beta x}\sin(\lambda_1(x-a))$ and $\Psi_0(x)=\sqrt{\frac{2}{b-a}}\,e^{-\beta x}\sin(\lambda_1(x-a))$ belong to $\hat\D$, and they satisfy the equation $\hat A\varphi_0=0$. Moreover, since $\hat B^\dagger=\sum_{n=0}^\infty\sqrt{\rho_{n+1}}\,\Psi_n\otimes\overline{\varphi_{n+1}}$, we have also $\hat B^\dagger\Psi_0=0$. Here, given $f,g,h\in\Hil_1$, we have defined $(h\otimes\overline{g})f:=\left<g,f\right>h$. Hence, Assumptions {\bf p1.} and {\bf p2.} are both satisfied. Also, we can check that $\hat A^\dagger=\sum_{n=1}^\infty\sqrt{\rho_{n}}\,\Psi_n\otimes\overline{\varphi_{n-1}}$, and a direct computation shows that also Assumption {\bf p3.} is satisfied. Finally, {\bf p4.} follows from our working assumption, in its stronger version (recall that $\F_\varphi$ and $\F_\Psi$ are Riesz bases). We also have
\be
\hat A^\dagger \hat B^\dagger\, \Psi_n=\rho_n\,\Psi_n,
\label{57}\en
as well as $\hat B^\dagger \hat A^\dagger\, \Psi_n=\rho_{n+1}\,\Psi_n$.

\vspace{2mm}

{\bf Remark:--} It should be observed that the stability of $\hat \D$ under the action of $\hat A^\sharp$ and $\hat B^\sharp$ is not evident. Therefore, the framework in Section \ref{sectnlpb} is not completely recovered. However, what we have discussed here shows that this stability is not really needed if we are just interested to obtain biorthogonal bases of two non self-adjoint operators, at least if we have (as we do have here) some reason which guarantees that all the vectors $\varphi_n$ and $\Psi_n$ belong to a same set, dense in the Hilbert space of the system. In other words, we can say that the Hamiltonian $H_{eff}$ considered in this section produces a slightly modified version of $\D$-NLPBs, sharing with those introduced in Section 2.2 most of their properties.

\vspace{2mm}

We continue our analysis introducing the operators $S_\varphi=\sum_{n=0}^\infty\varphi_n\otimes\overline{\varphi_n}$ and $S_\Psi=\sum_{n=0}^\infty\Psi_n\otimes\overline{\Psi_n}$, which are well defined everywhere in $\Hil_1$ since they can be written respectively as
\be
S_\varphi=\rho^{-2}=e^{2\beta x},\qquad S_\Psi=\rho^{2}=e^{-2\beta x}.
\label{en}\en
We see that they are self-adjoint, positive, and one the inverse of the other. Moreover, they are bounded and the following intertwining relations can explicitly be checked:
$$
S_\varphi (\hat B\hat A)^\dagger\Psi_n=(\hat B\hat A)S_\varphi\Psi_n,\quad S_\Psi (\hat B\hat A)\varphi_n=(\hat B\hat A)^\dagger S_\Psi\varphi_n,
$$
for all $n\geq0$. Similar intertwining relations can be found for the operator $\hat A\hat B$.

\vspace{2mm}

We end this section with the computation of the price kernels we have introduced in the previous section. In particular, repeating similar computations, here we find
$$
p_1(x,x';\tau)=\left<x,e^{-\tau H_{eff}}x'\right>=\frac{2}{b-a}e^{-\tau\gamma+\beta(x-x')}K(x,x';\tau),
$$
where
$$
K(x,x';\tau)=\sum_{n=0}^\infty e^{-\tau k^2 (n+1)^2}\sin\left[(n+1)\lambda_1(x-a)\right]\sin\left[(n+1)\lambda_1(x'-a)\right]
$$
and $k^2=\frac{\sigma^2\pi^2}{2(b-a)^2}$. A similar formula is found for $p_2(x,x';\tau)$, with $\beta$ replaced by $-\beta$. $K(x,x';\tau)$ can be rewritten in closed form in terms of the $\Theta$-function $\Theta_3(u,q)$, see \cite{grad}, Formula 8.180.4,
$$
\Theta_3(u,q)=1+2\sum_{m=1}^\infty q^{m^2}\cos(2mu),
$$
as follows:
$$
K(x,x';\tau)=\frac{1}{2}\left(K_1(x,x';\tau)-K_2(x,x';\tau)\right).
$$
Here
$$
K_1(x,x';\tau)=\frac{1}{2}\left[\Theta_3\left(\frac{\lambda_1(x-x')}{2},e^{-k^2\tau}\right)-1\right]
$$
and
$$
K_2(x,x';\tau)=\frac{1}{2}\left[\Theta_3\left(\frac{\lambda_1(x+x'-2a)}{2},e^{-k^2\tau}\right)-1\right].
$$

The pricing kernel could then be used, together with a suitable pay function $g(x')$, to recover the price of the options as in
$$
C_j(x;\tau)=\int_{\Bbb R} p_j(x,x';\tau) g(x'),
$$
$j=1,2$. We notice that, in the present settings we have, of course, two different prices $C_1$ and $C_2$, related one to the other simply by replacing $\beta$ with $-\beta$. In particular, they collapses when $\beta=0$, i.e. when $\sigma^2=2r$. More on the economical aspects of the model can be deduced following what is done in \cite{baaquie}. We will not consider these aspects here, since in this paper this is not our main interest.

\section{Conclusions}

We have discussed how to use the Black-Scholes equation for option pricing with constant volatility to produce examples of $\D$-PBs and $\D$-NLPBs. In particular we have found an example in which the eigenstates of the non self-adjoint Hamiltonians associated to the Black-Scholes equation form $\D$-quasi bases but not bases, and another example in which they form Riesz bases.  We have also deduced, for both these examples, the price kernels in closed forms.

As it is clear from our treatment, we have considered here mainly the mathematical aspects of the Schr\"odinger-like version of the Black-Scholes equation with just few comments on its economical meaning, and on the consequences of our approach. This is exactly what we hope to do next.

\section*{Acknowledgements}

The author acknowledges partial support from Palermo University and from G.N.F.M. of the INdAM.

\end{document}